\documentclass[aps,prb,twocolumn]{revtex4}
\usepackage{color}
\usepackage{bbm}
\usepackage{units}
\usepackage{graphicx}
\usepackage{float}
\usepackage{natbib}

\begin{document}

\title[Article Title]{Incorrect sample classification in ``Electron 
localization induced by intrinsic anion disorder in a transition metal 
oxynitride''}

\author{Arnulf M\"obius}

\email{arnulf.moebius@t-online.de}

\affiliation{Institute of Physics, University of Technology Chemnitz, 09107 Chemnitz, Germany}


\maketitle

\section{Introduction}

With their experimental study of temperature and composition dependences 
of transport properties of the disordered crystalline solid 
SrNbO$_{3 - x}$N$_x$, Ref.\ \onlinecite{Oka.etal.2021}, Daichi Oka and 
co-workers aimed to contribute to a better understanding of electron 
localization.  Interpreting their measured data according to the 
currently prevailing theories, they came across a gold nugget, but 
ignored it. That is, they obtained some data which clearly contradict 
these theories. Although those data are presented graphically in Ref.\ 
\onlinecite{Oka.etal.2021}, that serious inconsistency is not mentioned at all 
in its text. -- This is surprising since the authors cite a review and a 
comment, their Refs.\ 30 and 32 (here Refs.\ \onlinecite{Moebius.2019} and
\onlinecite{Moebius.1989}), to which discrepancies of this type 
are central. -- Those data are very interesting because they challenge 
the widely presumed continuity of the metal-insulator transition (MIT) 
in three-dimensional disordered solids, and thus the applicability of 
the very highly cited scaling theory of localization, Ref.\ 
\onlinecite{Abrahams.etal.1979}. In detail:

In identifying the MIT, Daichi Oka and co-workers rely on the 
theoretical ideas of Anderson localization 
\cite{Anderson.1958,Abrahams.etal.1979} and of the interference of 
electron-electron interaction and elastic impurity scattering of 
electrons \cite{Altshuler.Aronov.1979}. In particular, they do so when 
analysing the temperature dependence of the conductivity, $\sigma(T)$, 
of the sample with $x = 0.96$, referred to as sample 0.96 in the 
following. According to the $T \rightarrow 0$ extrapolation of a 
$\sigma = a + b \, T^{\, 1 / 2}$ fit \cite{Altshuler.Aronov.1979}, this 
sample is classified as metallic in Ref.\ \onlinecite{Oka.etal.2021}; thus the 
authors conclude the MIT to happen within the $x$ interval (0.96,1.02). 
That data evaluation, however, is invalid for several reasons as the 
present reconsideration from three different perspectives shows.

\boldmath
\section{The logarithmic derivative of $\sigma(T)$}
\unboldmath

Consider first the logarithmic derivative 
$w(T) = \mbox{d} \ln \sigma / \mbox{d} \ln T$. If measured data really 
obey $\sigma = a + b \, T^{\, p}$ with positive $a$, $b$, and $p$, then 
the slope of $w(T)$ is positive -- as claimed in the text of Ref.\ 
\onlinecite{Oka.etal.2021} to be the case for sample 0.96 at low $T$ -- and 
$w(T) \rightarrow 0$ as $T \rightarrow 0$. The validity of these 
statements on $w(T)$ should be verifiable by looking at Fig.\ 1d of 
Ref.\ \onlinecite{Oka.etal.2021}. Unfortunately, however, evaluating the 
low-$T$ part of the $w(T)$ data for sample 0.96 shown therein is 
complicated by the wide width of the $w$ range presented. Nevertheless, 
a thorough inspection of this diagram, enlarged by means of the PDF 
viewer, yields information on sample 0.96 that differs qualitatively 
from the above conclusions drawn from the ansatz 
$\sigma = a + b \, T^{\, p}$:
(i) With decreasing $T$, between about 8~K and the lowest $T$ value
considered, roughly 2~K, $w(T)$ no longer decreases as for
$25\ {\rm K} > T \gtrsim 11\ {\rm K}$, but even increases slightly; this 
finding contradicts the above mentioned claim of Ref.\ 
\onlinecite{Oka.etal.2021}.
(ii) In the whole $T$ range shown, $w(T) > 0$. 
Taking these two points together, one naively expects that $w(T)$ tends 
to some finite positive value $w_0$ as $T \rightarrow 0$, so that 
$\sigma(T)$ vanishes as $T^{\, w_0}$ -- or that $w(T)$ even diverges in 
this limit, so that $\sigma(T)$ vanishes in some exponential manner --. 
Therefore, the electric transport in sample 0.96 is most likely 
activated instead of metallic.

To shed more light on the contradiction between this conclusion and the 
interpretation in Ref.\ \onlinecite{Oka.etal.2021}, we digitized the
information on sample 0.96 given in Figs.\ 1b, 1c, and 1d of Ref.\ 
\onlinecite{Oka.etal.2021} by means of WinDIG 2.5. Furthermore, for 
comparison, data for the sample with $x = 1.02$, presented in Figs.\ 1c, 
1d and in Suppl.\ Fig.\ 5 of Ref.\ \onlinecite{Oka.etal.2021}, were read 
out. The latter sample is referred to as sample 1.02 from now on.

\begin{figure*} [t]
\includegraphics[width=0.92\linewidth]{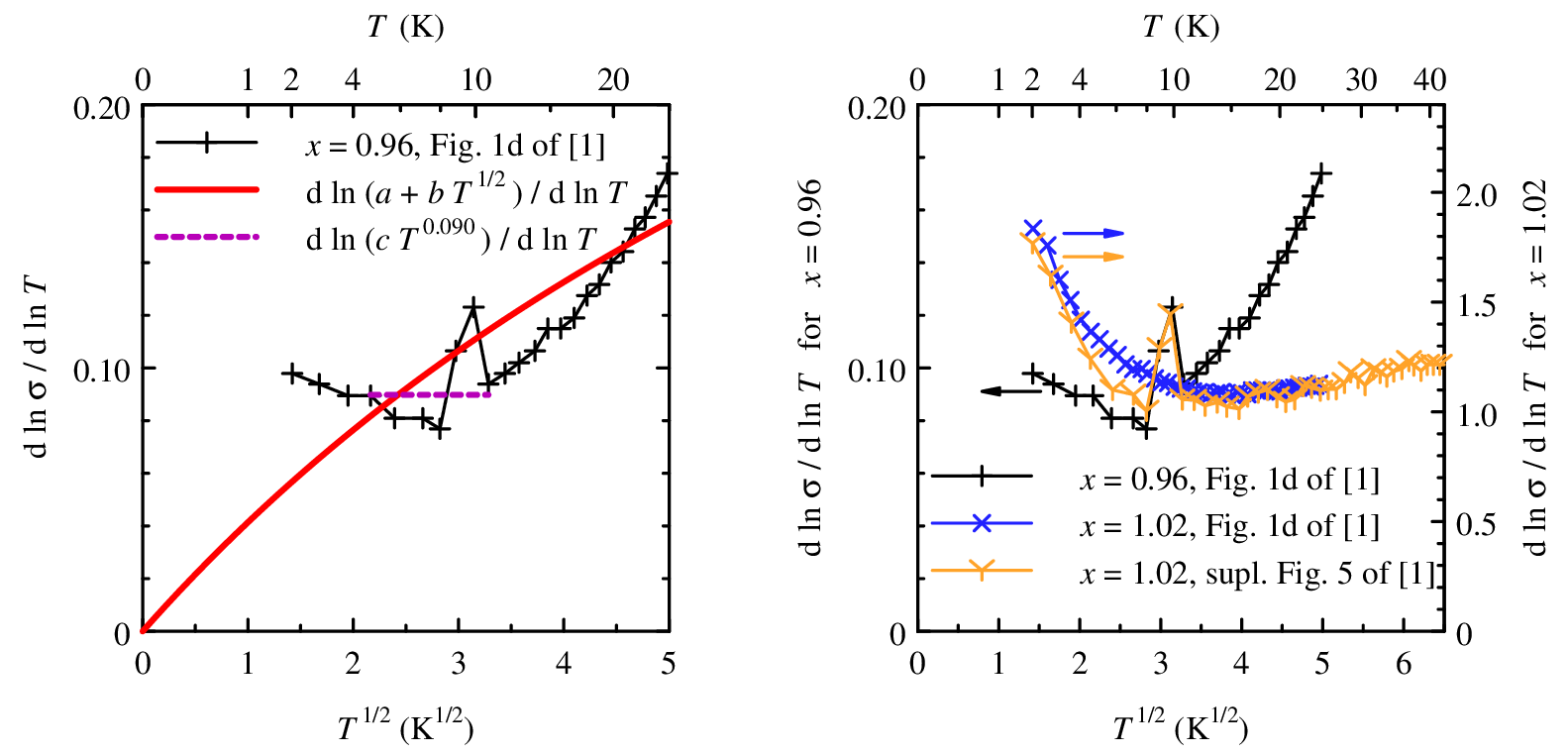}
\caption{Temperature dependence of the logarithmic derivative 
$w(T) = \mbox{d} \ln \sigma / \mbox{d} \ln T$ of the conductivity 
$\sigma(T)$ for the sample with $x = 0.96$, marked by black 
{\large $+$}. These data points were read out from Fig.\ 1d of Ref.\ 
\onlinecite{Oka.etal.2021}. In the left panel, they are compared to the 
logarithmic derivative of the fit function shown in Fig.\ 1c of 
Ref.\ \onlinecite{Oka.etal.2021}, given as red line here, and, additionally, 
to the logarithmic derivative of the power law interpolation between the 
$\sigma(4.7\ {\rm K})$ and $\sigma(10.7\ {\rm K})$ from Fig.\ 1c of 
Ref.\ \onlinecite{Oka.etal.2021}, shown as dashed magenta line. The right 
panel contrasts the $w(T)$ data points for the sample with $x = 0.96$ 
with $w(T)$ data points for the sample with $x = 1.02$,  where different 
$w$ scales (with equal zero) are used. The latter data points, read out 
from Fig.\ 1d of Ref.\ \onlinecite{Oka.etal.2021} and Suppl.\ Fig.\ 5 of Ref.\ 
\onlinecite{Oka.etal.2021}, are marked by blue {\large $\times$} and orange 
${\textsf Y}$, respectively; in the text, these data sets are referred 
to as set A and set B, respectively.}
\label{fig1}
\end{figure*}

The left panel of Fig.\ \ref{fig1} of our reconsideration makes the 
above described observations easily comprehensible. Therein, the $w(T)$ 
data points of sample 0.96 given in Fig.\ 1d of Ref.\ 
\onlinecite{Oka.etal.2021} are redrawn, where a roughly tenfold zoomed in $w$ 
scale is used. These data are contrasted with the analytically 
calculated logarithmic derivative of the 
$\sigma = a + b \, T^{\, 1 / 2}$ fit shown in Fig.\ 1c of Ref.\ 
\onlinecite{Oka.etal.2021}; for related details see below. Three 
characteristics of this set of redrawn data points catch the eye:
(1) When $T$ decreases from 25~K to about 10.5~K, $w(T)$ decreases too, 
though more and more slowly in this $w$ versus $T^{\, 1 / 2}$ 
representation. 
(2) Between about 10.5~K and about 4.7~K, $w(T)$ exhibits a peculiar 
feature, that is a pronounced hump and a subsequent swale. 
(3) Finally, below about 4.7~K, with further decreasing $T$, $w(T)$ 
slowly rises again; $w(T)$ increases by about 9~\% from 4.7~K to 2.0~K 
-- whereas the logarithmic derivative of the fit function (red curve) 
decreases by 31~\% over this interval --.

It is instructive to compare the behaviour of $w(T)$ for sample 0.96 to 
that of $w(T)$ for sample 1.02, nominally SrNbO$_2$N. For the latter 
sample, two data sets are available from Ref.\ \onlinecite{Oka.etal.2021}, 
that is from Fig.\ 1d and from Suppl.\ Fig.\ 5, in the following denoted 
as data sets A and B, respectively. Surprisingly, they differ 
considerably from each other, as is obvious upon first look at the 
respective $T$ regions around 10~K in both the diagrams. These two 
$w(T)$ data sets are reproduced in the right panel of our Fig.\ 
\ref{fig1}; they are contrasted with the $w(T)$ data for sample 0.96 
therein. 

According to the right panel of Fig.\ 1, in case of data set A, $w(T)$ 
smoothly runs through a minimum when $T$ decreases from 25~K to 2~K. 
This minimum is passed at about 15~K. Below it, $w(T)$ increases more 
and more rapidly with decreasing $T$. In contrast, when going from high 
to low $T$, the graph of set B exhibits a pronounced hump and a 
subsequent swale in the intermediate $T$ region, that is between 10.5~K 
and 4.7~K, similarly as $w(T)$ for sample 0.96.  Above 10.5~K and below 
4.7~K, however, the graphs of sets A and B roughly agree with each 
other. -- The small shift between these graphs at low $T$ would have 
been further diminished by ascribing the difference quotients to the 
respective interval centres, instead of to the lower interval boundaries 
as done in Ref.\ \onlinecite{Oka.etal.2021}. This way, by utilizing the 
symmetric difference quotient \cite{numdiff}, the numerical error would 
have been considerably reduced; see also Appendix C of Ref.\ 
\onlinecite{Moebius.2019}. --

There is a further inconsistency in Ref.\ \onlinecite{Oka.etal.2021}: the $T$ 
values of the $w(T)$ data points of sample 1.02 shown in its Fig.\ 1d 
(set A) are considerably closer together than the $T$ values of the 
$\sigma(T)$ data points of this sample presented in its Fig.\ 1c, 
although the $w_i$ should have been calculated from the $\sigma_i$ via 
difference quotients. -- The reader can easily check our statement by 
enlarging Figs.\ 1c and 1d of Ref.\ \onlinecite{Oka.etal.2021} by means of the 
PDF viewer. -- Remarkably, however, those $\sigma(T)$ data points in 
Fig.\ 1c relate to the same $T$ values as the $w(T)$ points for sample 
1.02 presented in Suppl.\ Fig.\ 5 of Ref.\ \onlinecite{Oka.etal.2021} (set B).

The most plausible hypothesis on the origin of the inconsistencies 
uncovered above is the following. Worried by the strange peculiarity 
obvious in Suppl.\ Fig.\ 5, Daichi Oka and co-workers may have 
remeasured $\sigma(T)$ for sample 1.02 following an improved procedure.
Unfortunately, however, they may have only updated Fig.\ 1d, letting
Fig.\ 1c and Suppl.\ Fig. 5 untouched. 

We now turn back to sample 0.96. To render the comparison of samples 
0.96 and 1.02 as informative as possible, we use different $w$ scales 
for the individual samples in the right panel of Fig.\ \ref{fig1}. The 
left scale is valid for sample 0.96; it agrees with the $w$ scale of 
the left panel of Fig.\ \ref{fig1}. The right scale applying to data for 
sample 1.02 has the same zero as the left one. Its unit, however, is 
chosen such that, at $T = 10.7\ {\rm K}$, the $w(T)$ data points for 
both the samples roughly fall together in this diagram.

Two striking similarities between the peculiar features of the $w(T)$ 
for sample 0.96 and for set B of sample 1.02 are clearly recognizable in
the right panel of Fig.\ \ref{fig1}:
(i) Both humps appear in the same $T$ region. 
(ii) Both humps have approximately the same shape and the same relative 
magnitude. 
These two findings, together with the absence of such a peculiar feature 
in the data set A for sample 1.02, suggest the following interpretation. 
Both the peculiar features are probably artifacts. Presumably, they 
originate from similar errors of the related $T$ values, caused by any 
shortcoming of the measuring procedure. Since Daichi Oka et al.\ 
apparently took values with slowly sliding temperature, insufficient 
equilibration together with a change of the cooling / heating rate may 
have been the origin of the peculiar features of the $w(T)$ for sample 
0.96 and for set B of sample 1.02. 

Together, the above discussed two comparisons in the right panel of 
Fig.\ 1 enable the following conclusion: the peculiar feature, that 
is hump plus swale in the intermediate $T$ range, does not disprove the 
validity of the $w(T)$ data points for $T \le 4.7\ {\rm K}$. With 
respect to sample 0.96, the trustworthiness of these low-$T$ data points 
is further supported by the logarithmic derivative of the power law 
interpolation between $\sigma(10.7\ {\rm K})$ and 
$\sigma(4.7\ {\rm K})$; this function provides a plausible estimate of 
$w(T)$ for the intermediate $T$ range, $w = 0.090$, see the left panel 
of Fig.\ \ref{fig1}. Therefore, it seems highly probable that, for 
sample 0.96, the slow increase of $w(T)$ with decreasing $T$ below about 
4.7~K is real, that it is not an artifact of temperature or resistivity 
measurements.

Apart from the mysterious feature hump plus swale, the graphical 
representations of the $w(T)$ data sets of sample 0.96 and of sample 
1.02 in our Fig.\ \ref{fig1} have qualitatively similar shapes -- though 
they differ in the position of the minimum and in the associated $w$ 
value --. This resemblance provides additional support for the above 
reclassification of sample 0.96. 

In this context, we stress that, apart from hump and swale, the 
appearance of both the $w(T)$ data sets of the samples 0.96 and 1.02 is 
rather typical for disordered solids, even if the minimum value of 
$w(T)$ is as small as roughly 0.1. The reader may compare them with the 
$w(T)$ data sets for compensated crystalline Si:(P,B) in Fig.\ 1 of 
Ref.\ \onlinecite{Hirsch.etal.1989}, an unusually unbiased and constructive 
reply to a comment. Furthermore, he/she may compare with data for 
crystalline Si:As and CdSe:In in Figs.\ 4 and 13, respectively, of the 
review Ref.\ \onlinecite{Moebius.2019} as well as with the ones for the 
amorphous solids a-Si$_{1 - x}$Cr$_x$ and a-Si$_{1 - x}$Mn$_x$ in Figs.\ 
7 and 8, respectively, of the review Ref.\ \onlinecite{Moebius.Adkins.1999}.
Correspondingly, having the four qualitative scenarios analysed in 
Section 5 of Ref.\ \onlinecite{Moebius.2019} in mind, under certain 
conditions, one can consider the $w(T)$ data of sample 0.96 as an 
indication of the existence of a finite minimum metallic conductivity.

Nevertheless, alternatively, it might be speculated that, for sample 
0.96, the negative slope of $w(T)$ below about 4.7~K could arise from 
only a part of the sample exhibiting metallic conduction, means from the 
sample being significantly inhomogeneous; compare Ref.\ 
\onlinecite{Hirsch.etal.1989}. 

However, due to a scaling law discovered roughly four decades ago 
\cite{Moebius.etal.1985}, it 
seems not unlikely that the observed negative slope of $w(T)$ for sample 
0.96 is a generic feature: for a-Si$_{1 - x}$Cr$_x$ films deposited by 
means of e-beam evaporation, the $T$ and $x$ dependences of $\sigma$ 
were observed to satisfy $\sigma(T,x) = \sigma_{\rm{sc}}(T / T_0(x))$ 
if 
$\sigma(T,x) < \sigma_0 = 270 \pm 50 \ {\rm \Omega}^{-1} {\rm cm}^{-1}$ 
and $100\ {\rm mK} < T < 50\ {\rm K}$. In 
words: under these conditions, $\sigma(T,x)$ is completely determined 
by the quotient of $T$ and an empirically introduced characteristic 
temperature $T_0(x)$. Since $\sigma_{\rm{sc}}(T / T_0)$ vanishes in an 
exponential manner as $T / T_0 \rightarrow 0$, this behaviour is related 
to activated conduction. -- Concerning exponential $T$ dependences, such 
scaling relations were observed to hold in several other homogeneous 
solids too \cite{Moebius.1985}. -- 

Note: that scaling law implies the existence of a finite minimum 
metallic conductivity, in contradiction to Ref.\ 
\onlinecite{Abrahams.etal.1979}. The scaling behaviour, however, breaks down 
in consequence of annealing, that is in consequence of the formation of 
nanocrystallites, see Subsection 3.3 of Ref.\ \onlinecite{Moebius.etal.1985}. 
Nevertheless, the appearance of $\sigma(T,x)$ varies only gradually in 
this process, not abruptly.   

When we have the possibility of such a scaling in mind, what do we have 
to look out for when studying other solids? For the reconsideration at 
hand particularly important: within the validity range of this scaling 
law, the slope of $w(T)$ is always negative. In various cases, this 
finding has motivated the study of $w(T)$ as cross-check of the 
interpretation of $\sigma(T)$ in terms of metallic conduction, see 
Refs.\ \onlinecite{Moebius.1989,Hirsch.etal.1989}, as well as the review Ref.\ 
\onlinecite{Moebius.2019} and references therein.

For the reconsideration at hand similarly important: the validity range 
of the scaling law mentioned above encompasses not only a conductivity 
region in which fast, exponential $T$ dependences are observed, but, 
furthermore, also a conductivity region where $\sigma(T)$ varies only 
slowly, in some non-exponential manner with $T$. 

This arises from the nature of the $x$ dependence of the characteristic 
temperature $T_0(x)$, which is defined by 
$\sigma(T,x) \propto \exp(-(T_0(x)/T)^{1/2})$ in the low-temperature 
limit. Apparently, $T_0(x) \rightarrow 0$ as $x$ tends to its critical 
value, at which the nature of conduction changes from activated to 
metallic, compare Fig.\ 3 of Ref.\ \onlinecite{Moebius.etal.1985}. This 
behaviour of $T_0(x)$ has a fundamental consequence: {\sl the closer the 
composition of the sample to that at the MIT, the higher is the lowest 
experimentally accessible temperature}. Our only at first glance absurd 
statement relates to the point that it is not the ratio of $T$ to the 
absolute unit Kelvin that is crucial here, but that it is the ratio of 
$T$ to the characteristic hopping temperature, $T_0(x)$: for 
$T \gg T_0(x)$, $\sigma(T/T_0(x))$ varies in some non-exponential 
manner, even if, for example, $T = 1\ {\rm K}$. 

Therefore, the following interpretation difficulty may arise in any 
experiment: prior to the transition from activated to metallic 
conduction, there is some finite $x$ range in which only rather flat, 
non-exponential $T$ dependences of $\sigma$ can be observed; compare 
the schematic diagram Fig.\ 2 of Ref.\ \onlinecite{Moebius.Adkins.1999}. 
Without a really unbiased, very thorough analysis -- as the evaluation 
of $w(T)$ --, such $\sigma(T)$ data may easily be misinterpreted to 
indicate metallic conduction.

\begin{figure*} [t]
\includegraphics[width=0.92\linewidth]{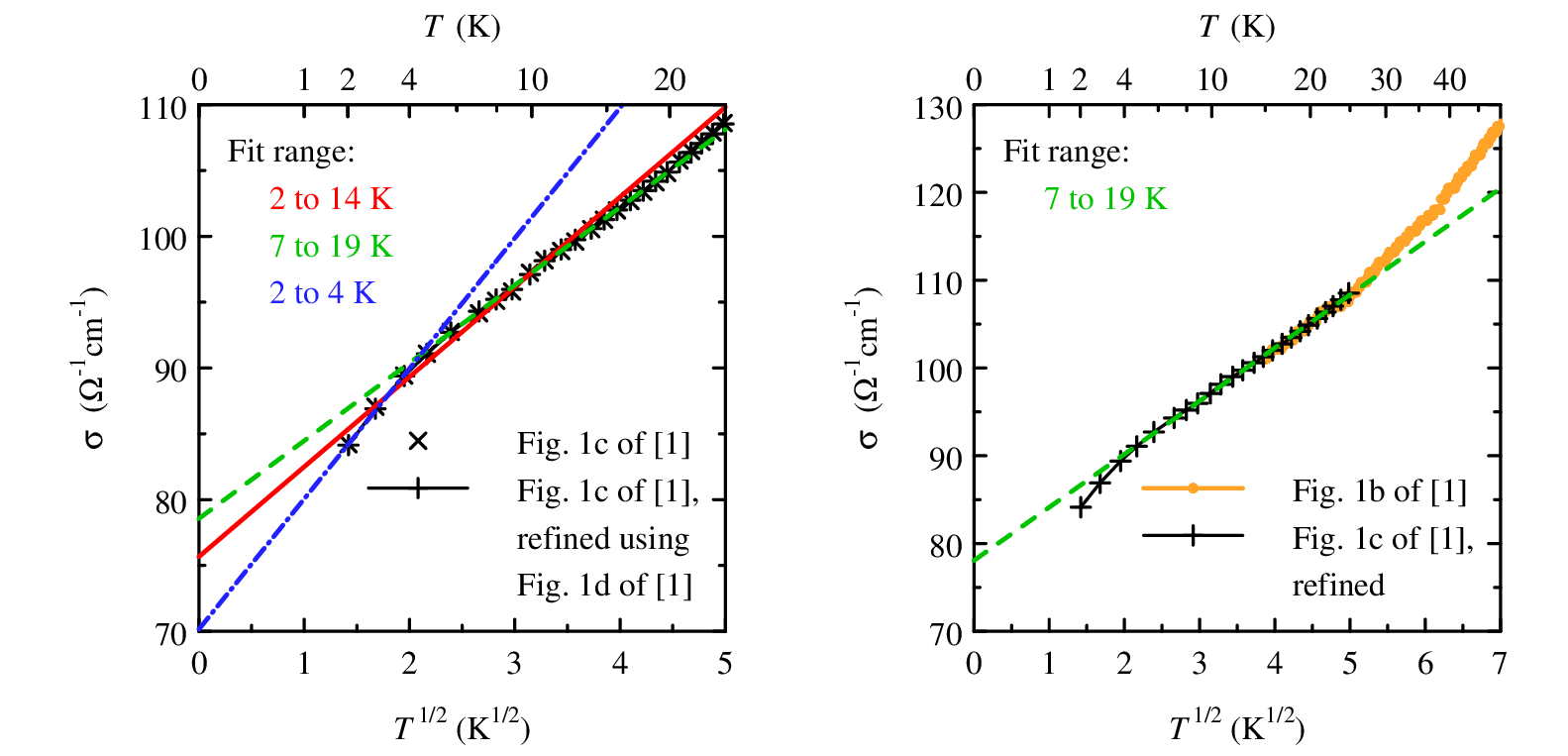}
\caption{Temperature dependence of the conductivity, $\sigma(T)$, for 
the sample with $x = 0.96$. According to Refs.\ 
\onlinecite{Oka.etal.2021,Altshuler.Aronov.1979}, it should be linear in such 
a $\sigma$ versus $T^{\, 1 / 2}$ plot. Data points read out directly 
from Fig.\ 1c of Ref.\ \onlinecite{Oka.etal.2021} are marked by 
{\large $\times$}, whereas the points which were refined additionally 
incorporating Fig.\ 1d of Ref.\ \onlinecite{Oka.etal.2021} -- see text -- 
are given as {\large $+$}. Left panel: Comparison of three fits of 
$\sigma = a + b \, T^{\, 1 / 2}$ to the refined data points focusing
on different temperature ranges: the solid red, dashed green, and dashed 
dotted blue lines relate to the intervals 2 to 14~K, 7 to 19~K, and 
2 to 4~K, respectively. Right panel: representation of the fit 
concerning the $T$ interval 7 to 19~K in an extended $T$ range. For 
that, additionally, conductivity data were read out from Fig.\ 1b of 
Ref.\ \onlinecite{Oka.etal.2021} by means of automatic pixel by pixel 
digitization.} 
\label{fig2}
\end{figure*}

\boldmath
\section{Approximation by $\sigma(T) = a + b \, T^{\, 1 / 2}$}
\unboldmath

Assume now, a reader is not aware of this principal difficulty and has 
no information on $w(T)$. Then Fig.\ 1c of Ref.\ \onlinecite{Oka.etal.2021}, 
in particular the seemingly rather good quality of the 
$\sigma = a + b \, T^{\, 1 / 2}$ fit to the data of sample 0.96, may 
mislead him/her: it could make him/her feel sure about this sample being 
clearly metallic. So the following questions arise. Is the quality of 
the approximation shown in Fig.\ 1c of Ref.\ \onlinecite{Oka.etal.2021} really 
as convincing as it seems at first glance? How reliable are the 
parameter estimates of this approximation? Why does it work quite well 
over a rather wide $T$ range? 

Insight into these interpretation issues is gained by studying under 
which conditions significant deviations of the measured data points of 
sample 0.96 from the adjusted function $\sigma = a + b \, T^{\, 1 / 2}$ 
occur. For such an analysis, one needs as precise as possible 
$(T_i,\sigma_i)$ data. Thus, first, we checked the precision of the data 
read out from Fig.\ 1c of Ref.\ \onlinecite{Oka.etal.2021} by comparing the 
$w_i$ values calculated from the digitized $(T_i,\sigma_i)$ to the $w_i$ 
data read out from Fig.\ 1d of Ref.\ \onlinecite{Oka.etal.2021}. Then, we used 
the digitized $T_i$ and $w_i$ to refine the $\sigma_i$ values. This 
precision improvement bases on the digitized $w_i$ having a smaller 
discretization error than the differences between neighbouring 
$\sigma_i$ as read out. Additionally, a tiny constant shift of all $w_i$ 
was allowed to minimize the sum of the squares of the small corrections 
of the $\sigma_i$ values in the refinement; the optimum constant shift 
corresponds to roughly half a pixel in Fig.\ 1d of Ref.\ 
\onlinecite{Oka.etal.2021}.

The results are presented in our Fig.\ \ref{fig2}. Its left panel 
displays $\sigma$ vs.\ $T^{\, 1 / 2}$ with much better $\sigma$ 
resolution than Fig.\ 1c of Ref.\ \onlinecite{Oka.etal.2021}. Two data sets 
are shown therein: the set of $\sigma(T)$ points obtained directly by 
means of digitization of Fig.\ 1c of Ref.\ \onlinecite{Oka.etal.2021}, as well 
as the refined set obtained as described in the previous paragraph. Both 
these data sets deviate only slightly from each other. Completing, the 
right panel of Fig.\ \ref{fig2} presents an overview over a considerably 
wider $T$ range, from 0~K to 49~K, though with somewhat lower $\sigma$ 
resolution. It combines the refined set and data read out from Fig.\ 1b 
of Ref.\ \onlinecite{Oka.etal.2021}.

\begin{figure*} [t]
\includegraphics[width=0.92\linewidth]{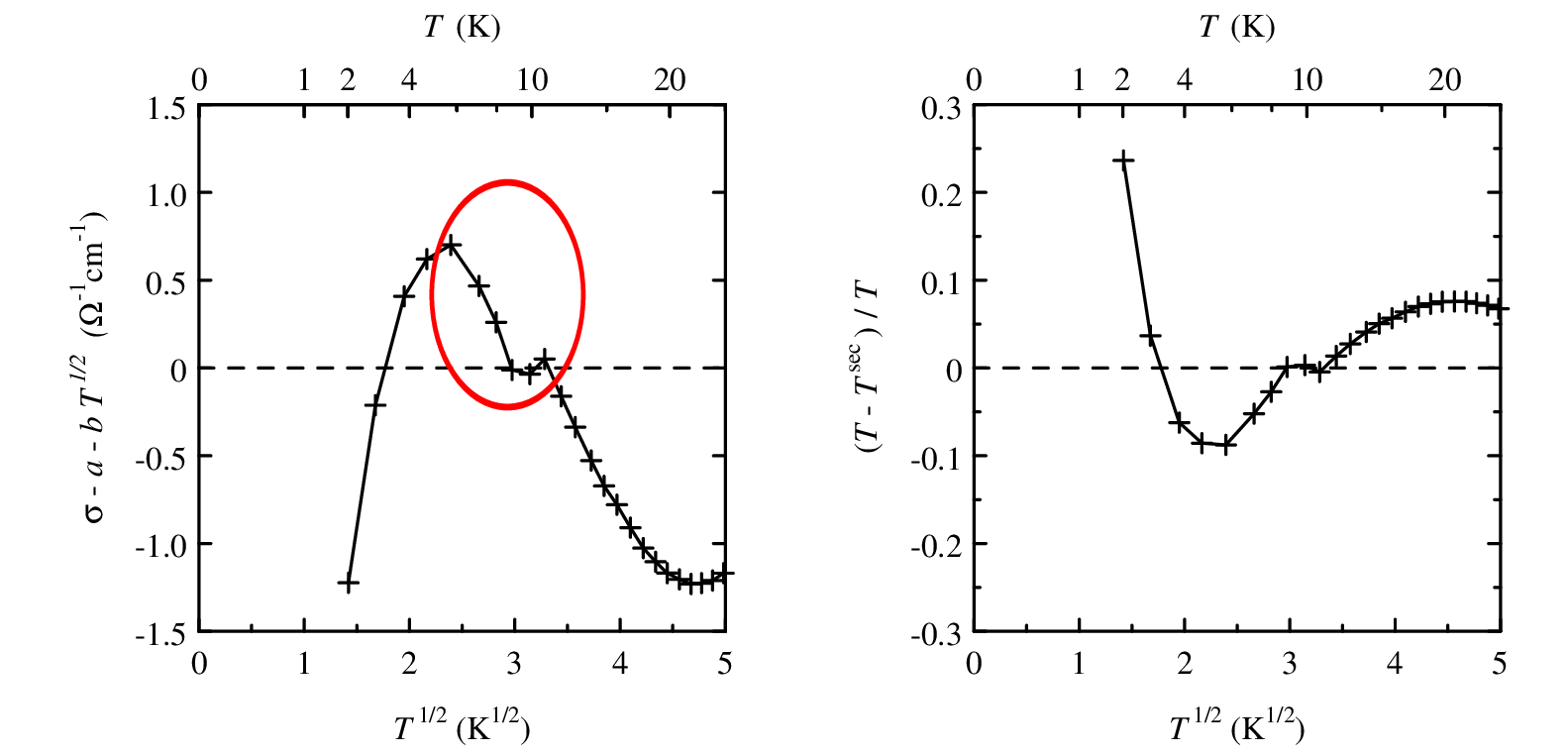}
\caption{Temperature dependences of the deviations between measured 
data and their $\sigma = a + b \, T^{\, 1 / 2}$ approximation, relating 
to the interval $2\ \rm{K} < T < 14\ \rm{K}$, for the sample with 
$x = 0.96$. The left panel presents the absolute conductivity misfit, 
while the right panel shows the relative temperature misfit, see text.
} 
\label{fig3}
\end{figure*}

The left panel of Fig.\ \ref{fig2} compares three fits of the function
$\sigma = a + b \, T^{\, 1 / 2}$ to different subsets of the set of the 
refined digitized data: 
(1) The solid red line corresponds to the approximation shown in Fig.\ 
1c of Ref.\ \onlinecite{Oka.etal.2021}: the parameters $a$ and $b$ are 
adjusted to optimally reproduce the measured points within the $T$ 
interval 2~K to 14~K. This fit yields 
$\sigma(0\ {\rm K}) = a = 75.7\ {\rm \Omega}^{-1} {\rm cm}^{-1}$, in 
very good agreement with the value read out from Fig.\ 1c of Ref.\ 
\onlinecite{Oka.etal.2021}, $75.8\ {\rm \Omega}^{-1} {\rm cm}^{-1}$. 
\linebreak
(2) The dashed green line refers to a fit to the data points within an 
intermediate $T$ region, that is $7\ \rm{K} < T < 19\ \rm{K}$. 
According to this approximation, 
$\sigma(0\ {\rm K}) = 78.6\ {\rm \Omega}^{-1} {\rm cm}^{-1}$. 
(3) The dashed-dotted blue line represents a fit to the data points 
for the three lowest $T$ values, $2\ \rm{K} < T < 4\ \rm{K}$; it 
yields $\sigma(0\ {\rm K}) = 70.2\ {\rm \Omega}^{-1} {\rm cm}^{-1}$.

Three conclusions can be drawn from this comparison. First, substantial 
deviations from the $\sigma = a + b \, T^{\, 1 / 2}$ approximation are 
clearly recognizable. Second, these deviations have primarily a 
systematic, non-random character: apparently, the experimental 
$\sigma(T^{\, 1 / 2})$ crosses the solid red line only twice. Third, the 
result of the extrapolation to $T = 0 \ \rm{K}$ depends considerably on 
which $T$ interval is taken into account in the fit: the lower the upper 
boundary of this $T$ interval, the smaller is the estimated value of 
$\sigma(0\ {\rm K})$.

In spite of the above demonstrated low-temperature extrapolation 
uncertainty, the fit of $\sigma = a + b \, T^{\, 1 / 2}$ to the measured 
$\sigma(T)$ data given in Fig.\ 1c of Ref.\ \onlinecite{Oka.etal.2021} seems 
to yield a quite reasonable approximation within a rather wide interval. 
The question is why. Inclusion of data of sample 0.96 from Fig.\ 1b of 
Ref.\ \onlinecite{Oka.etal.2021} into the comparison solves this puzzle: in 
the right panel of our Fig.\ \ref{fig2}, the set of refined data of 
sample 0.96 is supplemented with data from Fig.\ 1b of Ref.\ 
\onlinecite{Oka.etal.2021} for $15 \ \rm{K} \lesssim T < 49 \ \rm{K}$. 
Contrasting this composed graph with the 
$\sigma = a + b \, T^{\, 1 / 2}$ fit to all refined data points within 
the range $7\ \rm{K} < T < 19\ \rm{K}$, marked by the dashed green line, 
suggests the following explanation. 

Overall, the experimental relation $\sigma(T^{\, 1 / 2})$ shown in 
the right panel of Fig.\ \ref{fig2} has an inverse S shape and, 
correspondingly, an inflection point within the considered $T$ range; 
this characteristic temperature amounts to (very) roughly 10~K. So the 
dashed green line represents a linear fit focusing on the surroundings 
of an inflection point. Therefore, the quadratic term of the Taylor 
expansion of $\sigma(T^{\, 1 / 2})$ around the fit interval centre is 
almost negligible so that the region of approximately linear behaviour 
is comparably wide. Thus, simply for numerical reasons, it is not 
surprising that the dashed green line provides a quite reasonable 
description over the rather wide $T$ range from 4~K to 25~K. 

As cross-check of this interpretation, consider the fit focusing on the 
interval 2~K to 4~K -- dashed-dotted blue line in the left panel of 
Fig.\ \ref{fig2} --. It concerns a region considerably away from the 
inflection point. As expected, it seems to have a far smaller 
applicability range; in this case, the judgement bases on the estimation 
of the upper half width of this range.

\section{Sample as secondary thermometer}

Now, we have a closer look at the deviations of the $\sigma(T)$ data 
points of sample 0.96 from their $\sigma = a + b \, T^{\, 1 / 2}$ 
approximation. The left panel of our Fig.\ \ref{fig3} shows the 
differences between the refined $\sigma_i$ data and the corresponding 
values $a + b \, T_i^{\, 1 / 2}$ of the fit for the region 2~K to 
14~K; this approximation focuses on the same $T$ interval as considered
in Ref.\ \onlinecite{Oka.etal.2021}. -- Note: the resolution of the vertical 
axis of the left panel of Fig.\ \ref{fig3} is by a factor 13 higher than 
that of the $\sigma$ axis of the left panel of Fig.\ \ref{fig2}. -- 

Loosely speaking, the left panel of Fig.\ \ref{fig3} presumes the $T$ 
values to be exact and shows how the deviation of the $\sigma$ value 
from its estimate depends on $T^{\, 1 / 2}$. Complementing this 
presentation, the right panel of Fig.\ \ref{fig3} relates to the 
opposite perspective: the $\sigma$ values are presumed to be exact, and 
we ask for the deviation of the measured $T$ value from the $T$ value 
estimated on the basis of the same $\sigma = a + b \, T^{\, 1 / 2}$ 
approximation. In other words, in the right panel of Fig.\ \ref{fig3}, 
the sample is considered as secondary thermometer: to each value of 
$\sigma_i$, a secondary temperature value, $T^{\rm sec}_i$, is 
ascribed. The corresponding relative misfit, 
$(T_i - T^{\rm sec}_i) / T_i$, is depicted versus the square root of 
the actual temperature, $T_i^{\, 1 / 2}$, in this diagram.

The left panel of Fig.\ \ref{fig3} confirms the presence of substantial 
systematic, non-random deviations between measured data and fit. 
Furthermore, it provides the following two interesting messages. (i) The 
very systematic $T$ dependence of the deviation between measured data 
and fit present above 10~K testifies the rather high precision of the 
resistivity measurements by Daichi Oka and co-workers. (ii) 
Simultaneously, however, this $T$ dependence exhibits a strange feature 
between about 5~K and about 10~K, marked by a red ellipse; that 
presumable inconsistency correlates with the peculiar feature hump plus 
swale illuminated in our analysis of the logarithmic derivative of 
$\sigma(T)$ in Section 2. Both these statements are confirmed by the 
right panel of Fig.\ \ref{fig3}.

Finally, we emphasize two quantitative conclusions which can be drawn 
from the right panel of Fig.\ \ref{fig3}. First, above 3~K, the relative 
misfit, \linebreak $(T_i - T^{\rm sec}_i) / T_i$, varies between as much 
as -9~\% and +8~\%. Second, below 3~K, where the relative misfit is 
positive, it increases rapidly with decreasing $T$ and reaches a value 
of 24~\% at 2~K! Of course, this sharp increase should be checked by 
additional measurements at intermediate $T$ values. Furthermore, an 
extension of the considered $T$ range down to 1.8~K -- presumably in 
reach of the equipment used in Ref.\ \onlinecite{Oka.etal.2021} -- should be 
very helpful in clarifying the situation.  

Both these findings are further strong arguments against the 
significance of the $T \rightarrow 0$ extrapolation of the 
$a + b \, T^{\, 1 / 2}$ fit to the measured $\sigma(T)$ data. This 
extrapolation, however, is very important to Ref.\ 
\onlinecite{Oka.etal.2021}.

\section{Conclusions}

We summarize: regarding the sample with $x = 0.96$, Daichi Oka and 
co-workers comprehensively reported on their quite careful, albeit not 
problem-free determinations of $\sigma(T)$ and $w(T)$ in Figs.\ 1b, 1c, 
and 1d of Ref.\ \onlinecite{Oka.etal.2021}. In doing so, unfortunately, they 
used inappropriate $\sigma$ and $w$ scales. Here, zooming into Figs.\ 1c 
and 1d of Ref.\ \onlinecite{Oka.etal.2021}, we have uncovered a fundamental 
contradiction between the authors' theoretical interpretation of these 
$\sigma(T)$ measurements, on the one hand, and their $w(T)$ data, on the
other hand. The authors claim in their text that, for $x = 0.96$, the 
$w(T)$ dependence, determined in an ansatz-free manner, would
have a posive slope at low temperatures so that this sample would 
be metallic. In fact, however, below 4.7~K, $w(T)$ does not decrease 
with decreasing $T$, but even increases slightly, indicating activated 
transport. 

Motivated by this contradiction, we have checked in detail the quality 
of the $\sigma = a + b \, T^{\, 1 / 2}$ fit for the sample with 
$x = 0.96$ shown in Fig.\ 1c of Ref.\ \onlinecite{Oka.etal.2021}. In our Fig.\ 
2, the uncertainty of corresponding $T \rightarrow 0$ extrapolations is 
demonstrated. Simultaneously, it is shown, that it is the vicinity of an 
inflection point of $\sigma(T^{1/2})$ that pretends a high significance 
of this approximation. Finally, focusing on the deviations between this 
fit and the measured data, we have visualized, in our Fig.\ 3, the 
doubtfulness of this approximation of the data points taken at the 
lowest temperatures.

It is surprising that the contradiction highlighted above was not 
mentioned at all in the text of Ref.\ \onlinecite{Oka.etal.2021}. It was not 
mentioned though, concerning the interpretation of $w(T)$ data, the 
authors cite Ref.\ \onlinecite{Moebius.2019}; that recent critical review 
discusses observations and consequences of similar behaviour of $w(T)$ 
for various other disordered solids in great detail. 

Thus an unusual question suggests itself: to what extent is the 
narrative that the scaling theory of localization 
\cite{Abrahams.etal.1979} qualitatively correctly describes the MIT of 
various three-dimensional disordered systems a result of a strong 
confirmation bias? Serious doubts about the applicability of this theory 
have been on the table since the identification of a scaling law for the 
temperature dependence of the conductivity in the hopping region of 
various three-dimensional disordered systems
\cite{Moebius.1985,Moebius.etal.1985} roughly four decades ago. This 
empirical finding implies the existence of a finite minimum metallic 
conductivity, which in contrast is denied by the scaling theory of 
localization \cite{Abrahams.etal.1979}. The mentioned doubts were 
strongly enhanced by the surprising discovery of a metallic phase 
in two-dimensional disordered systems by Sergey Kravchenko and 
co-workers \cite{Kravchenko.etal.1994,Kravchenko.Sarachik.2004} in 1994: 
the existence of such a phase is ruled out by the scaling theory of 
localization \cite{Abrahams.etal.1979}.

Hence, further thorough and unbiased experimental studies of the MIT in 
disordered solids are urgently needed. So it will certainly be highly 
valuable if Daichi Oka and co-workers repeat the $\sigma(T)$ measurement 
of their sample with $x = 0.96$ with increased precision -- provided 
this sample is sufficiently long-term stable --. In doing so, compared 
to Ref.\ \onlinecite{Oka.etal.2021}, a greater part of the data should be 
taken close to the lower end of the $T$ region considered: constant 
quotient of subsequent $T$ values is appropriate. Precision and 
significance of such measurements will benefit from carefully 
thermalizing the setup of sample and thermometer before each resistivity 
measurement, instead of measuring with slowly sliding temperature. 
Including into the proposed study additional samples with slightly 
different compositions will presumably be fruitful.

\vspace{0.5cm}

\begin{center} {\bf Acknowledgments} \end{center}

Various critical remarks by Manuel Richter were very helpful in 
improving the manuscript.

\begin{center} {\bf Ethics declaration} \end{center}

The author declares no competing interests.

\begin{center} {\bf Data availability} \end{center}

All data obtained in this analysis are available from the author upon 
reasonable request.

\end{document}